\documentclass[a4paper,twoside]{article}
\usepackage{amssymb}
\usepackage{amsmath}
\usepackage{amsfonts}
\usepackage{graphicx}

\begin{document}

\title{\Large{\bf 
Evaporating loop quantum black hole 
}}
\author{\\ Leonardo Modesto\footnote{Electronic address: modesto@bo.infn.it or leonardo.modesto@virgilio.it}
 \\[1mm]
\em\small{ Department of Physics, Bologna University \& INFN Bologna,}\\[-1mm]
\em\small{V. Irnerio 46, I-40126 Bologna, EU.}\\[-1mm] \em\small{Centre de Physique
 Th\'eorique de Luminy, 
 Universit\'e de la M\'editerran\'ee,
 }\\[-1mm] \em\small{ 
 Case 907, F-13288
 Marseille, EU.}
   }

\date{\ } 
\maketitle

\begin{abstract}
In this paper we 
obtain the black hole metric from  
a semiclassical analysis of loop quantum black hole.
Our solution and the Schwarzschild one tend to match well 
at large distances from 
Planck region. In $r=0$ the semiclassical metric is regular
and singularity free in contrast to the classical one.
By using the new metric we calculate the Hawking temperature and the
entropy. For the entropy we obtain the logarithmic correction
to the classical area law. 
Finally we study the mass evaporation process and we show
the mass and temperature tend to zero at infinitive time.

\end{abstract}

\section*{Introduction}

Quantum gravity is the theory by which we tray to reconcile general relativity and quantum
mechanics. 
In general relativity the space-time is dynamical then 
it is not possible to study other interactions on a fixed background because
the background itself is a dynamical field.
 The theory called
 ``loop quantum gravity" (LQG) \cite{book}
is the most widespread nowadays.
This is one of the non perturbative and 
background independent approaches to quantum gravity
(another non perturbative approach to quantum gravity is called  ``asymptotic safety 
quantum gravity" \cite{MR}). 
Loop quantum gravity is a quantum geometric 
fundamental theory that reconciles general relativity 
and quantum mechanics at the Planck scale.
The main problem nowadays is to connect this fundamental theory
with standard model of particle physics and in particular with the 
effective quantum field theory.
In the last two years great progresses has been done to
connect  (LQG) with the low energy physics by
the general boundary approach \cite{MV}, \cite{ModestoRovelli}.
Using this formalism it has been possible 
to calculate the graviton propagator in four \cite{Rovelli1}, \cite{BMSR}
and three dimensions \cite{Simone1}, \cite{Simone2}.
In three dimensions it has been showed that a noncommuative 
field theory can be obtained from spinfoam models 
\cite{Freidel}. 
Similar efforts in four dimension are going in progress \cite{FB}. 

Early universe and black holes 
are other interesting places for testing the validity of LQG.
In the past years 
applications of LQG ideas to minisuperspace models 
lead to some interesting results in those fields. 
In particular it has been 
showed in cosmology \cite{Boj}, \cite{MAT} and recently in black hole physics
\cite{work1}, \cite{work2}, \cite{work3}, \cite{ABM} that it is 
possible to solve the cosmological singularity problem and the
black hole singularity problem by using 
tools and ideas developed in full loop quantum gravity theory. 
 
We can summarize this short introduction to ``loop quantum gravity
program" in two research lines; the first one dedicated to obtain 
quantum field theory from the fundamental theory and the other one
dedicated to apply LQG to cosmology and astrophysical 
objects where extreme energy conditions need to
know a quantum gravity theory.

In this paper we concentrate our attention on the second research line.
We study the black hole physics 
using ideas 
suggested by loop quantum gravity at the semiclassical level
(by ``loop quantum black hole" \cite{ABM}, 
extending the metric of obtained \cite{SS}
to all space-time).
The new metric is regular in $r=0$ where the classical singularity is localized 
and we are interesting to calculate the temperature, entropy
and to analyze the evaporation process.

This paper is organized in two section as follows.  In the first section we briefly
recall the semiclassical Schwarzschild solution inside the black hole
 \cite{SS} and we extend the solution outside the event horizon
 showing the regularity of the curvature invariant $\forall \,  r\geqslant 0$.
In the second section we calculate the Hawking temperature 
and the entropy in terms of the event horizon area. In the same section
we study also the mass evaporation process 
discussing the new physics suggested by loop quantum gravity.

\section{Semiclassical black hole solution 
}
In this section we summarize the solution calculated in paper 
\cite{SS} and we extend the solution to all the space-time.
We start to study the region inside the event 
horizon where the Ashtekar's connection and density triad 
are 
\begin{eqnarray}
&& A= c \tau_3 d x + b \tau_2 d \theta - b \tau_1 \sin \theta d \phi + \tau_3 \cos \theta d \phi,
\nonumber \\
&&E = p_c \tau_3 \sin \theta \frac{\partial}{\partial x} + p_b \tau_2 \sin \theta \frac{\partial}{\partial \theta} - p_b \tau_1 \frac{\partial}{\partial \phi},
\label{AE}
\end{eqnarray}
(where $\tau_i = - \frac{i}{2} \sigma_i$ and $\sigma_i$ are the Pauli matrices).
The variables in the phase space 
are: $(b, p_b)$, $(c, p_c)$,
and the Poisson algebra is:
$\{c, p_c \} = 2 \gamma G_N$, $\{b, p_b \} = \gamma G_N$.
The Hamiltonian constraint 
of  ``loop quantum
black hole" \cite{ABM} in terms of 
holonomies 
\footnote{
The Hamiltonian constraint in terms of holonomies is 
\begin{eqnarray}
&& \mathcal{C}^{\delta} =  - \frac{2 \hbar N}{\gamma^3 \delta^3 l_P^2}
{\rm Tr} \left(\sum_{ijk} \epsilon^{ijk} h_i^{(\delta)} h_j^{(\delta)} h_i^{(\delta) -1} h_k^{(\delta)}
\left\{h_k^{(\delta) -1}, V \right\} + 2 \gamma^2 \delta^2 \tau_3 h_1^{(\delta)}\left\{h_1^{(\delta) -1}, V\right\} \right) = \nonumber \\
&& \hspace{0.4cm} = - \frac{8 \pi N}{\gamma^2 \delta^2} 
\left\{ 2 \sin \delta b \ \sin \delta c \ \sqrt{|p_c|} +
(\sin^2 \delta b + \gamma^2 \delta^2) \frac{p_b \ \mbox{sgn}(p_c)}{\sqrt{|p_c|}} \right\},
\label{CH}
\end{eqnarray} 
where the holonomies in the directions $r, \theta, \phi$, integrated along curves of 
length $\delta$, are 
\begin{eqnarray}
h_1 = \cos \frac{\delta c}{2} + 2 \tau_3 \sin \frac{\delta c}{2}, \hspace{1cm}
h_2 = \cos \frac{\delta b}{2} - 2 \tau_1 \sin \frac{\delta b}{2}, \hspace{1cm}
h_3 = \cos \frac{\delta b}{2} + 2 \tau_2 \sin \frac{\delta b}{2},
\end{eqnarray}
and $V = 4 \pi \sqrt{|p_c|} p_b$ is the spatial section volume.
}
depends explicitly on the parameter $\delta$ that 
defines the length of the curves along which we integrate the connections.
The parameter $\delta$ is not an external cutoff but 
instead a result of full loop quantum gravity \cite{LoopOld}.
The Hamiltonian constraint $\mathcal{C}^{\delta}$ in (\ref{CH})
can be substantially simplified   
in the particular gauge 
$
N = \frac{\gamma \sqrt{|p_c|} \mbox{sgn}(p_c) \delta^2}{16 \pi G_N \sin \delta b}$
\begin{eqnarray}
\mathcal{C}^{\delta} = - \frac{1}{2 \gamma G_N} 
\Big\{ 2 \sin \delta c  \  p_c +
\Big(\sin \delta b + \frac{\gamma^2 \delta^2}{\sin \delta b} \Big) 
p_b \Big\}.
\label{FixN}
\end{eqnarray}
From (\ref{FixN}) we obtain two independent sets of equations of motion on the 
phase space 
\begin{eqnarray}
&& \dot{c} = - 2 \sin \delta c, \hspace{3cm} \dot{p_c} = 2 \delta p_c \cos \delta c 
\nonumber \\
&&\dot{b} = - \Big(\sin \delta b + \frac{ \gamma^2 \delta^2}{\sin \delta b} \Big), \hspace{1.5cm}
\dot{p_b} =  \delta \, \cos \delta b \Big(1 - \frac{ \gamma^2 \delta^2}{\sin^2 \delta b} \Big) p_b.
\end{eqnarray}  
Solving the first three equations and using the Hamiltonian constraint $\mathcal{C}^{\delta} =0$
we obtain \cite{SS}
\begin{eqnarray}
 && c(t) =  \frac{2}{\delta} \arctan \Big( \mp \frac{\gamma \delta m p_b^{(0)}}{2 t^2}  \Big),
\nonumber \\
&& p_c (t) = \pm 
\frac{1}{t^2} 
 \Big[\Big(\frac{\gamma \delta m p_b^{(0)}}{2}\Big)^2  + t^4 \Big] \nonumber \\,
&& \cos \delta b = \sqrt{1 + \gamma^2 \delta^2} 
\left[ \frac{\sqrt{1 + \gamma^2 \delta^2} + 1 - \Big(\frac{ 2 m}{t} \Big)^{\sqrt{1 + \gamma^2 \delta^2}}
(\sqrt{1 + \gamma^2 \delta^2} - 1)}
{\sqrt{1 + \gamma^2 \delta^2} + 1 + \Big(\frac{ 2 m}{t} \Big)^{\sqrt{1 + \gamma^2 \delta^2}}
(\sqrt{1 + \gamma^2 \delta^2} - 1)}
\right], \nonumber \\
&& p_b(t) = -  \frac{2 \ \sin \delta c \
\sin \delta b \ p_c }{\sin^2 \delta b + \gamma^2 \delta^2},
\label{Sol.cpcbpb}
\end{eqnarray}
(we have used the parametrization $t \equiv e^{\delta T}$ \cite{SS}).  
Respect to the classical Schwarzschild solution 
$p_c$ has an absolute minimum in $t_{min} = (\gamma \delta m p_b^{(0)}/2)^{1/2}$,
and $p_{c}(t_{min}) = \gamma \delta m p_b^{(0)}>0$.
The solution presents an inner horizon in \cite{SS}   
\begin{eqnarray}
t^{\ast} = 2 m \left(\frac{\sqrt{1 + \gamma^2 \delta^2} - 1}{\sqrt{1 + \gamma^2 \delta^2} + 1}\right)^{\frac{2}{\sqrt{1 + \gamma^2 \delta^2}}}.
\end{eqnarray}
\begin{figure}
 \begin{center}
  \includegraphics[height=6cm]{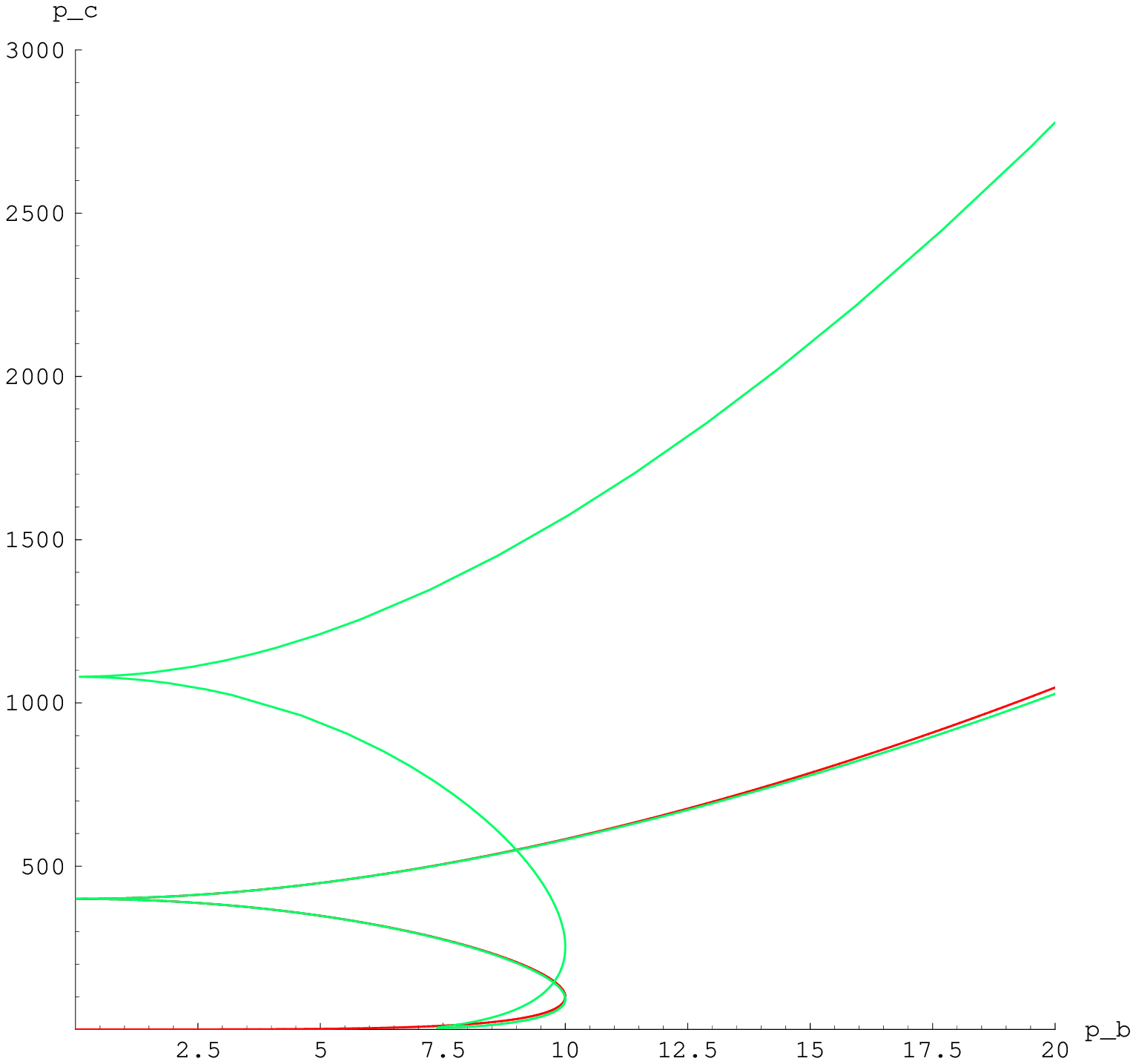}
  \includegraphics[height=6cm]{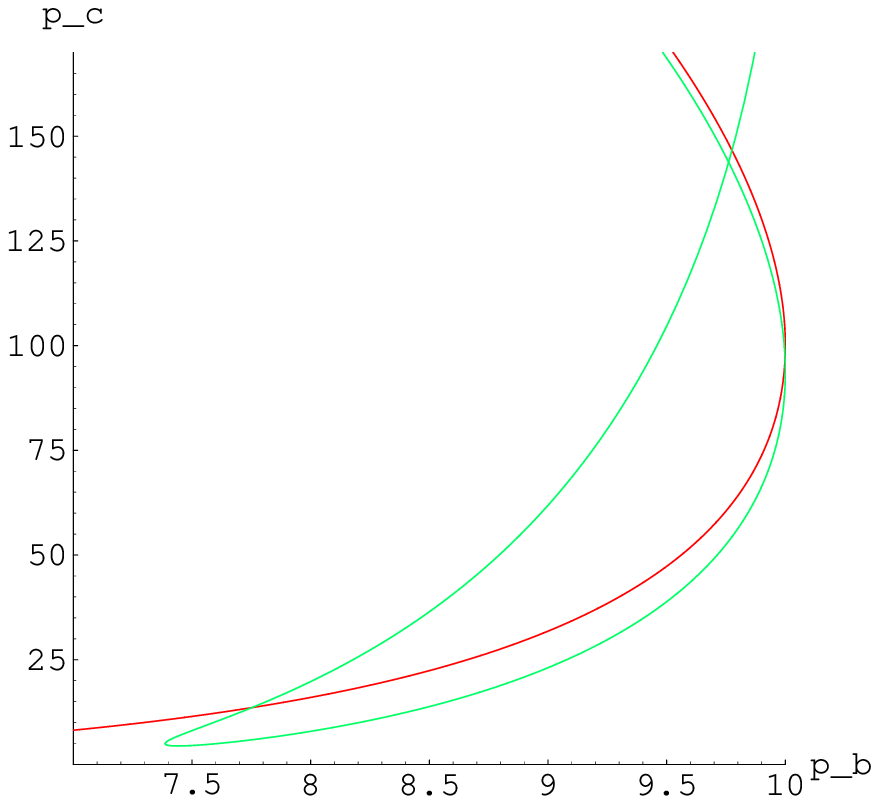}
  \end{center}
  \caption{\label{pbpc} 
   Semiclassical dynamical trajectory in the plane $p_b - p_c$.
   The plots for $p_c > 0$ and for $p_c < 0$ are disconnected and symmetric
   but we plot only the positive values of $p_c$.  
  The red trajectory corresponds to
                 the classical Schwarzschild solution and the 
                 green trajectory corresponds to the semiclassical solution. 
                The plot on the right represents a zoom of trajectory region 
                where the semiclassical analysis becomes relevant.}
  \end{figure}

We study the trajectory in
the plane $p_c-p_b$ and we compare the result with
the Schwarzschild solution.
In Fig.\ref{pbpc}  we have a parametric plot of $p_c$ and $p_b$ for $m = 10$ and $\gamma \delta \sim 1$ to amplify the quantum gravity effects in the plot (in \cite{SS} it has been showed that $\delta \sim 10^{-33}$
\footnote{The parameter $\delta$ is related to the minimum area eigenvalue that in
quantum geometry is $ \sim l_P^2$ \cite{LoopOld}.
}
; if we introduce a characteristic size length $L$ for the system under 
consideration 
we can set $\delta = l_P/L$). 
In Fig.\ref{pbpc} we can follow the trajectory 
from $t > 2m$ where the classical (red trajectory) and the semiclassical (green trajectory) 
solution are very close. For $t = 2m$,
$p_c \rightarrow (2m)^2$ and $p_b \rightarrow 0$ (this point 
corresponds to the Schwarzschild radius).
From this point decreasing $t$ we reach 
a minimum value for $p_{c,m} \equiv p_c(t_{min}) >0$. From $t=t_{ min}$, $p_c$
starts to grow again until $p_b=0$, this point corresponds to a new horizon 
in $t = t^{\ast}$ localized. In the time interval $t < t_{min}$, $p_c$ grows
together with $|p_b|$ 
and the functions $p_c, |p_b| \rightarrow \infty$ for $t \rightarrow 0$;
in particular $|p_b| \sim t^{- \sqrt{1+ \gamma^2 \delta^2}}$ for $t \sim 0$.


\paragraph{Metric form of the solution.}
In this paragraph we present the metric form of the solution.
The Kantowski-Sachs metric is 
$ds^2 = - N^2(t) dt^2 + X^2(t) dr^2 + Y^2(t)(d \theta^2 + \sin \theta d \phi^2)$
and the metric components are related to the connection variables by 
\begin{eqnarray}
Y^2(t) =|p_c(t)|, \,\,\,\,\,\,\, X^2(t) = \frac{p_b^2(t)}{|p_c(t)|}, \,\,\,\,\,\,\, 
N^2(t) = \frac{\gamma^2 \delta^2 |p_c(t)|}{ t^2 \sin^2 \delta b}. 
\label{metric}
\end{eqnarray}
\begin{figure}
 \begin{center}
  \includegraphics[height=5.5cm]{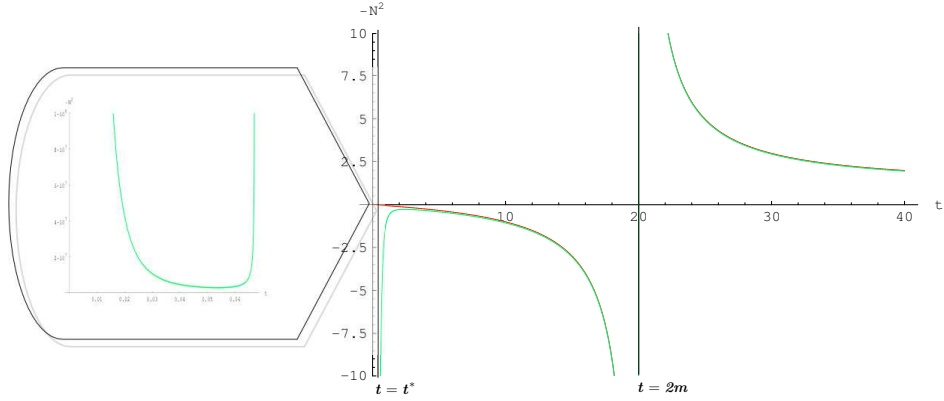}
  \end{center}
  \caption{\label{N2} Plot of the lapse function $ - N^2(t)$ 
 for $m = 10$ and $\gamma \delta \sim 1$
 (in the horizontal axis 
  we have the temporal coordinate $t$ and in the vertical axis the lapse function). 
                 The red trajectory corresponds to
                 the classical Schwarzschild solution inside the event horizon and the 
                 green trajectory corresponds to the semiclassical solution.
                 In the left side we have a zoom of $- N^2(t)$
                  in the region $0 \leqslant t \leqslant t^{\ast}$.}
  \end{figure}
The explicit form of the lapse function $N(t)^2$ in terms of the temporal coordinate $t$ is 
\begin{eqnarray}
N^2(t) = \frac{\gamma^2 \delta^2 \left[ \left(\frac{\gamma \delta m}{2 t^2}\right)^2 +1 \right]}{
  1 - (1 + \gamma^2 \delta^2)
\left[ \frac{\sqrt{1 + \gamma^2 \delta^2} + 1 - \left(\frac{2m}{t} \right)^{\sqrt{1 + \gamma^2 \delta^2}}
(\sqrt{1 + \gamma^2 \delta^2} - 1)}
{\sqrt{1 + \gamma^2 \delta^2} + 1 + \left(\frac{2m}{t} \right)^{\sqrt{1 + \gamma^2 \delta^2}}
(\sqrt{1 + \gamma^2 \delta^2} - 1)}
\right]^2   
}.
\label{N2metric}
\end{eqnarray}
In Fig.\ref{N2} we have a plot of the lapse function $-N(t)^2$ ($\forall t \geqslant 0$), 
for $m = 10$ and $\gamma \delta \sim 1$
(we have taken $\gamma \delta \sim 1$ to amplify, in the plot, the loop quantum gravity 
modifications at the Planck scale).
The red trajectory corresponds to the classical solution $ -1/(2m/t -1)$ and the green 
line to the semiclassical solution. We can observe the two solutions are identically in the
space-time region far from the Planck scale. In particular they have the same 
asymptotic limit for $t \gg 2m$. In the region $0 \leqslant t \leqslant t^{\ast}$ a plot of 
$- N^2(t)$ is given in the square on the left side in Fig.\ref{N2}. 

Using the second relation of (\ref{metric}) we can obtain also the other components of the
metric \cite{SS}, 
\begin{eqnarray}
X^2(t) = \frac{(2 \gamma \delta m)^2 \  
                         \left(1 - (1 + \gamma^2 \delta^2)
\left[ \frac{\sqrt{1 + \gamma^2 \delta^2} + 1 - \left(\frac{2m}{t} \right)^{\sqrt{1 + \gamma^2 \delta^2}}
(\sqrt{1 + \gamma^2 \delta^2} - 1)}
{\sqrt{1 + \gamma^2 \delta^2} + 1 + \left(\frac{2m}{t} \right)^{\sqrt{1 + \gamma^2 \delta^2}}
(\sqrt{1 + \gamma^2 \delta^2} - 1) }
\right]^2   \right) \  t^2 }{
(1 + \gamma^2 \delta^2)^2 \left(1 -
\left[ \frac{\sqrt{1 + \gamma^2 \delta^2} + 1 - \left(\frac{2m}{t} \right)^{\sqrt{1 + \gamma^2 \delta^2}}
(\sqrt{1 + \gamma^2 \delta^2} - 1)}
{\sqrt{1 + \gamma^2 \delta^2} + 1 + \left(\frac{2m}{t} \right)^{\sqrt{1 + \gamma^2 \delta^2}}
(\sqrt{1 + \gamma^2 \delta^2} - 1)}
\right]^2\right)^2
\Big[\Big(\frac{\gamma \delta m}{2}\Big)^2  + t^4 \Big]
}.
\label{X2}
\end{eqnarray}
and the radius of the $S^2$ sphere as a function of the temporal coordinate is 
\begin{eqnarray}
Y^2(t) =  
\frac{1}{t^2} 
 \Big[\Big(\frac{\gamma \delta m}{2}\Big)^2  + t^4 \Big].
 \label{Y}
\end{eqnarray}
\begin{figure}
 \begin{center}
  \includegraphics[height=4cm]{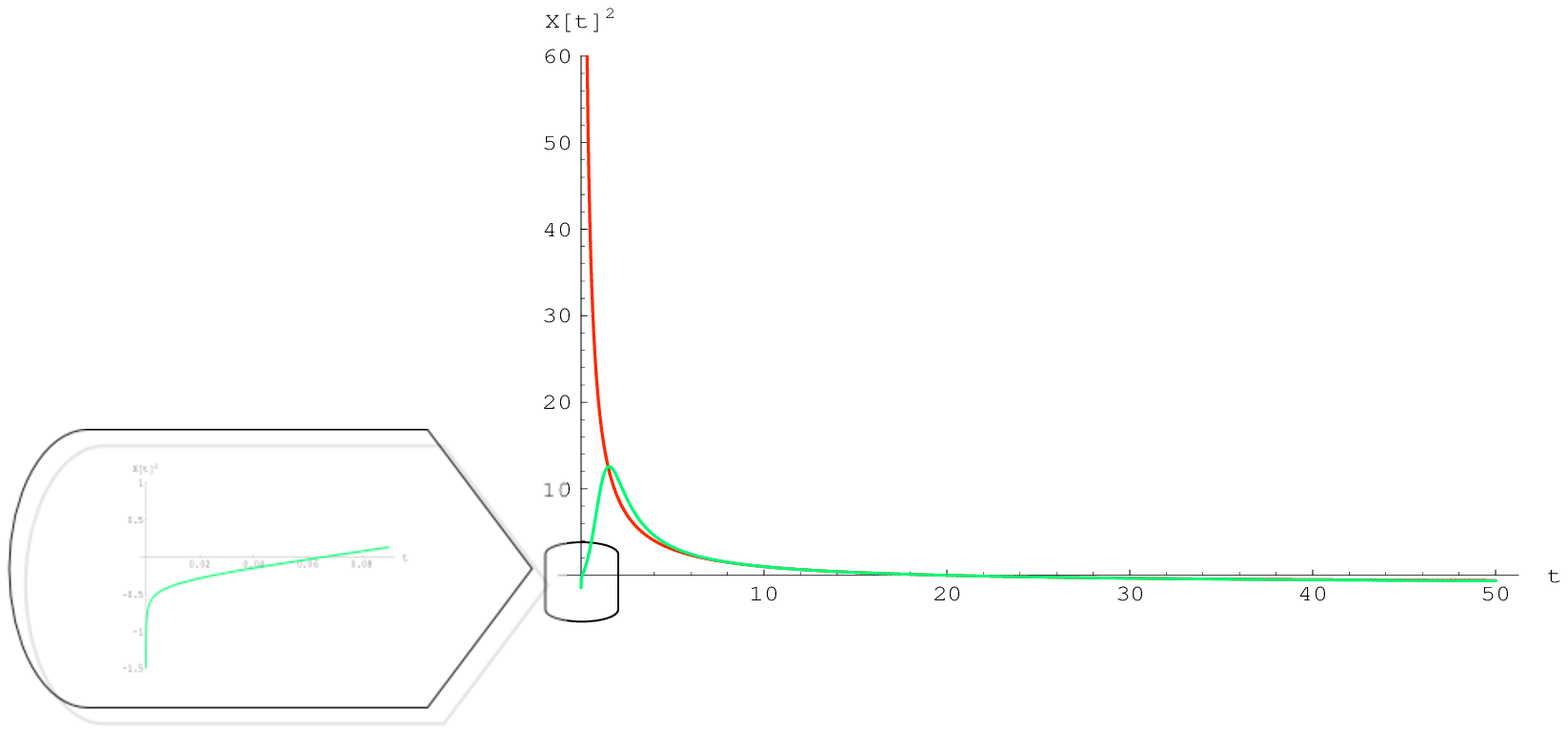}
  \includegraphics[height=4cm]{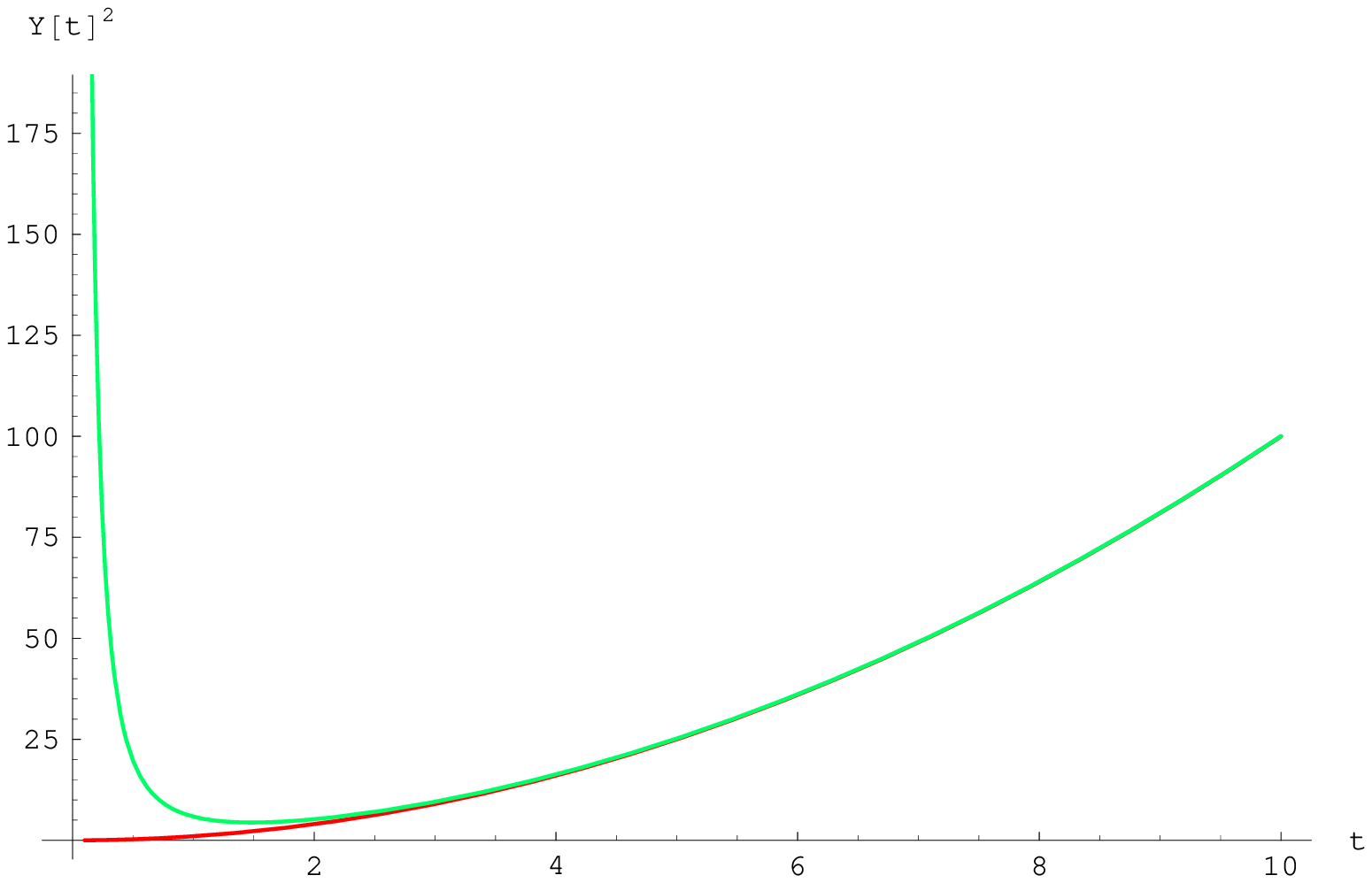}
  \end{center}
  \caption{\label{X2All} 
   Plot of $X^2(t)$ and $Y^2(t)$ 
 for $m = 10$, $\gamma \delta \sim 1$ and $\forall t \geqslant 0$; the event horizon 
 for this particular values of the parameters is in $t = 20$. 
                 The red trajectory corresponds to
                 the classical Schwarzschild solution and the 
                 green trajectory corresponds to the semiclassical solution.}
  \end{figure}
  
\noindent In Fig.\ref{X2All} a plot of $X^2(t)$ inside and outside
the event horizon, $\forall t \geqslant 0$ is represented. From the solution (\ref{X2})
we can calculate the limit for $t \rightarrow 0$ and for $t \rightarrow \infty$ and we 
obtain $X^2(t \rightarrow 0) \rightarrow - \infty$ 
(
$X^2(t) \sim  t^{- \gamma^2 \delta^2}$ for $t \sim 0$) 
and $X^2(t \rightarrow \infty) \rightarrow -1$
(the large distance limit is $X^2(t) \sim -1 + \gamma^2 \delta^2 \ln \frac{2 m}{t}$,
but the logarithmic correction is meaningful for a distance $t \sim e^{1/\delta^2}$ which
is longer the universe radius).
The value of the coordinate $t$ where the solution $X^2(t)$ changes sign 
corresponds to the inner horizon in $t = t^{\ast}$ localized. 
In the second picture of Fig.\ref{X2All} we have a plot of $Y^2(t)$ and we can note a substantial 
difference with the classical solution. In the classical case the $S^2$ sphere
goes to zero for $t \rightarrow 0$. In the semiclassical solution instead
the $S^2$ sphare bounces on a minimum value of the radius,
which is $Y^2(t_{min}) = \gamma \delta m$,
and it expands again to infinity for $t \rightarrow 0$.
The minimum of $Y^2(t)$ corresponds to the time coordinate 
$t_{min} = (m \gamma \delta/2)^{1/2}$.

It is useful make a change in variables from $t \rightarrow r$ 
in order to study the evaporation process,
and we focus 
our attention on the outside event horizon region toward the near horizon region.
From the outside event horizon point of view the causal structure of the space-time
is defined by the identifications $ - N^2(t) \rightarrow g_{rr}(r)$ and 
$X^2(t) \rightarrow g_{tt}(r)$.  We redefine also the $S^2$ sphere 
radius in terms of metric components, 
$Y^2(t) \rightarrow Y^2(r) = g_{\theta \theta}(r) = g_{\phi \phi}/\sin^2 \theta$. 
The solution can be summarized in the following table. 
\begin{center}
\begin{tabular}{|r|r|r|}
\hline
$ g_{\mu \nu} $& $\rm Semiclassical$&$ \rm Classical$\\
\hline
\hline
$g_{tt}(r)$ & 
 $\frac{(2 \gamma \delta m)^2 \  
                         \left(1 - (1 + \gamma^2 \delta^2)
\left[ \frac{\sqrt{1 + \gamma^2 \delta^2} + 1 - \left(\frac{2m}{r} \right)^{\sqrt{1 + \gamma^2 \delta^2}}
(\sqrt{1 + \gamma^2 \delta^2} - 1)}
{\sqrt{1 + \gamma^2 \delta^2} + 1 + \left(\frac{2m}{r} \right)^{\sqrt{1 + \gamma^2 \delta^2}}
(\sqrt{1 + \gamma^2 \delta^2} - 1) }
\right]^2   \right) }{
(1 + \gamma^2 \delta^2)^2 \left(1 -
\left[ \frac{\sqrt{1 + \gamma^2 \delta^2} + 1 - \left(\frac{2m}{r} \right)^{\sqrt{1 + \gamma^2 \delta^2}}
(\sqrt{1 + \gamma^2 \delta^2} - 1)}
{\sqrt{1 + \gamma^2 \delta^2} + 1 + \left(\frac{2m}{r} \right)^{\sqrt{1 + \gamma^2 \delta^2}}
(\sqrt{1 + \gamma^2 \delta^2} - 1)}
\right]^2\right)^2
\Big[\Big(\frac{\gamma \delta m}{2 r}\Big)^2  + r^2 \Big]
}$ & $ - (1 - \frac{2m}{r})$\\
\hline
\hline
$g_{rr}(r)$ & 
 $ - \frac{\gamma^2 \delta^2 \left[ \left(\frac{\gamma \delta m}{2 r^2}\right)^2 +1 \right]}{
  1 - (1 + \gamma^2 \delta^2)
\left[ \frac{\sqrt{1 + \gamma^2 \delta^2} + 1 - \left(\frac{2m}{r} \right)^{\sqrt{1 + \gamma^2 \delta^2}}
(\sqrt{1 + \gamma^2 \delta^2} - 1)}
{\sqrt{1 + \gamma^2 \delta^2} + 1 + \left(\frac{2m}{r} \right)^{\sqrt{1 + \gamma^2 \delta^2}}
(\sqrt{1 + \gamma^2 \delta^2} - 1)}
\right]^2   
}
$
  & $\frac{1}{1 - \frac{2m}{r}}$\\
\hline
\hline
$g_{\theta \theta}(r) = \frac{g_{\phi \phi}(r)}{\sin^2 \theta}$ & $\left(\frac{\gamma \delta m }{2 r}\right)^2  + r^2$ & $r^2$\\
\hline
\end{tabular}
\end{center}
If we develop the metric in the table by the parameter $\delta$ 
(or $l_P$ in dimensionless units) we obtain 
the Schwarzschild solution to zero order: $g_{tt}(r) = - (1 - 2m/r) + O(\delta^2)$, 
$g_{rr}(r) =  1/(1 - 2m/r) + O(\delta^2)$ and $g_{\theta \theta}(r) = g_{\phi \phi}(r)/\sin^2 \theta = r^2 + O(\delta^2)$.

\begin{figure}
 \begin{center}
  \includegraphics[height=6cm]{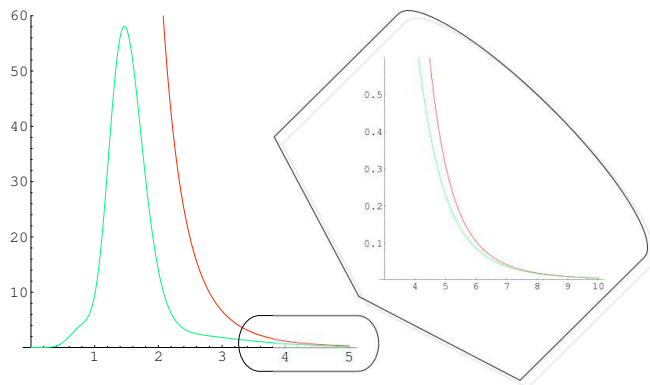}
  \end{center}
  \caption{\label{R2} 
   Plot of the invariant $\rm R_{\mu \nu \rho \sigma} \rm R^{\mu \nu \rho \sigma} $ 
 for $m = 10$, $\gamma \delta \sim 1$ and $\forall t \geqslant 0$; the large 
 $t$ behaviour 
  is $1/t^6$ as shown in the zoom on the right side.}
  \end{figure}

\paragraph{Regular semiclassical solution in $r = 0$.}
In this paragraph we show the semiclassical solution is regular in $r=0$ where
the classical singularity is localized. We calculate the curvature invariant 
$\rm R_{\mu \nu \rho \sigma} \rm R^{\mu \nu \rho \sigma}$ and we plot the result 
in terms of the variable $t$. 
 The classical singularity is in $t=0$ localized. 
 We express the curvature invariant in terms of the
functions $N(t)$, $X(t)$ and $Y(t)$ of the previous section and we obtain
\begin{eqnarray}
&& {\rm R_{\mu \nu \rho \sigma} \rm R^{\mu \nu \rho \sigma}} = 4 \Bigg[
\left(\frac{1}{ X N} \frac{d}{d t} \left( \frac{1}{N} \frac{d X}{dt} \right)\right)^2 + 
2 \left(\frac{1}{ Y N} \frac{d}{d t} \left( \frac{1}{N} \frac{d Y}{dt} \right)\right)^2+\nonumber \\
&& \hspace{2.3cm} + 2\left(   \frac{1}{ X N} \frac{d X}{d t} \, \frac{1}{ Y N} \frac{d Y}{d t} \right)^2 +
\frac{1}{Y^4 N^4}\left(N^2 + \Big(\frac{d Y}{d t}\Big)^2 \right)^2 \Bigg].
\label{CUR}
\end{eqnarray}
Introducing the  
explicit form of the metric in (\ref{CUR})
we obtain a regular quantity in $t=0$.
We give in Fig.\ref{R2} a plot of the result in term of the coordinate $t$.
From the plot it is evident that the curvature invariant tends to zero for 
$t\rightarrow 0$ and match with the classical quantity for large value of the 
time coordinate $t$. We give below the first correction in $\delta$ to
the curvature invariant 
\begin{eqnarray}
&& \hspace{-1.5cm} \rm R_{\mu \nu \rho \sigma} \rm R^{\mu \nu \rho \sigma} = 
\frac{48 m^2}{t^6}  \nonumber\\
&&\hspace{-0.5cm} + \frac{8 m \gamma^2 \delta^2}{t^{10}} 
\left(-33 m^3 + 12 m^2 t - 6 m^2 t^3 + 2 t^5 + 2m t^4 (\ln(8) -1) + 6 m t^4 \ln\big(\frac{m}{t} \big) \right) +O(\delta^3).
\label{R2App}
\end{eqnarray}
  From (\ref{R2App}) we can see that the first correction to the curvature invariant 
  is singular in $t=0$ and this is the same for each other orders. 
  This note shows that the regularity of the semiclassical solution is 
  a genuinely non perturbative result.
(For the semiclassical solution the trace of the Ricci tensor
($\rm R = R^{\mu}_{\mu}$) is not 
identically zero as for the Schwarzschild solution. We have calculated the 
trace invariant and we have showed that also this quantity is regular in $r =0$).
\begin{figure}
 \begin{center}
  \includegraphics[height=6cm]{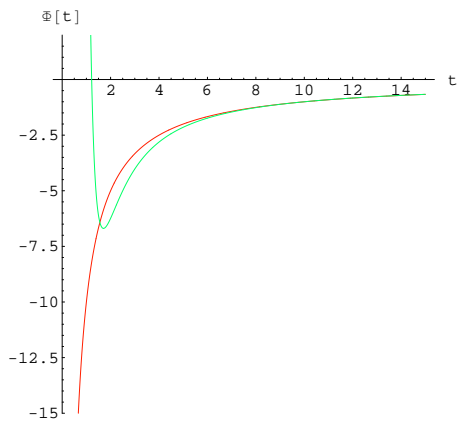}
  \end{center}
  \caption{\label{Pot} 
 Plot of 
 the first correction to the gravitational potential for 
 $m = 10$ and $\gamma \delta \sim 1$ to amplify the quantum gravity effects; 
 the red trajectory corresponds to the classical potential
 and the green trajectory to the semiclassical one.}
  \end{figure}

  \paragraph{Corrections to the Newtonian potential.}
  Another quantity that we can extract from the semiclassical metric is 
  the first correction of gravitational potential.
  The gravitational potential is related to the metric by 
  $\Phi(r) = - \frac{1}{2} (g_{tt}(r) + 1)$.
  Developing  
  the metric component (\ref{X2}) by the parameter $\delta$
  we obtain the first correction to the gravitational potential 
  \begin{eqnarray}
  \Phi(r) = - \frac{m}{r} - \frac{\gamma^2 \delta^2}{2} \left(1 + \ln\left(\frac{2 m}{r} \right) 
  - \frac{m}{r} \ln\left(\frac{2 m}{r} \right) - \frac{3 m}{r} 
  + \frac{2 m^2}{r^2} + \frac{m^2}{4 r^4} - \frac{m^3}{2 r^5}
  \right).
  \label{poten}
  \end{eqnarray}
  The parameter $\delta = l_P/L$, where $L$ is the characteristic length 
  of the physical system, plays the role of dimensionless Plank length.
  When we restore the length units the first five terms are multiplied by 
  $(l_P/L)^2$ and the last two by $l_P^2$.

\section{Temperature, entropy and evaporation}
The form of the metric calculated in the preview section
and in \cite{SS} has the general form 
\begin{eqnarray}
ds^2 = - g(r) dt^2 + f^{-1}(r) dr^2 + h^2(r) (d \theta^2 + \sin^2 \theta d \phi^2),
\label{generalmet}
\end{eqnarray}
where the functions $f(r)$, $g(r)$ and $h(r)$ depend on the mass parameter $m$
and are given in the table of the first section.
\begin{figure}
 \begin{center}
  \includegraphics[height=3cm]{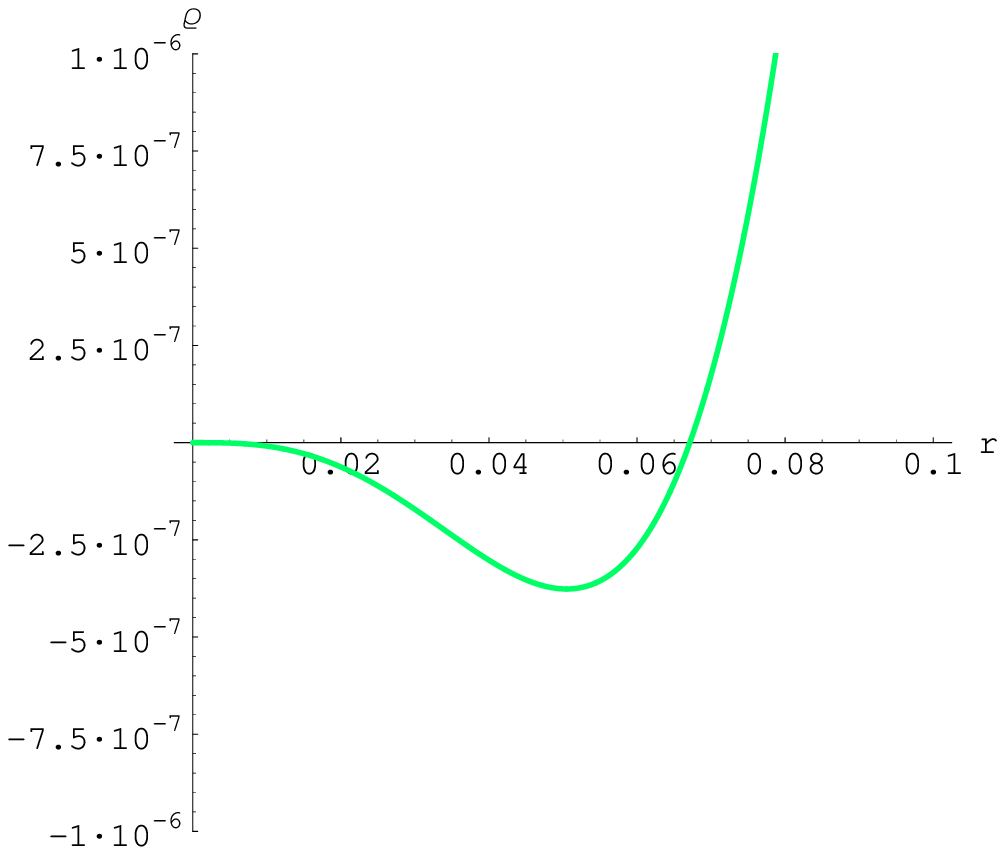} \hspace{1cm}
  \includegraphics[height=3cm]{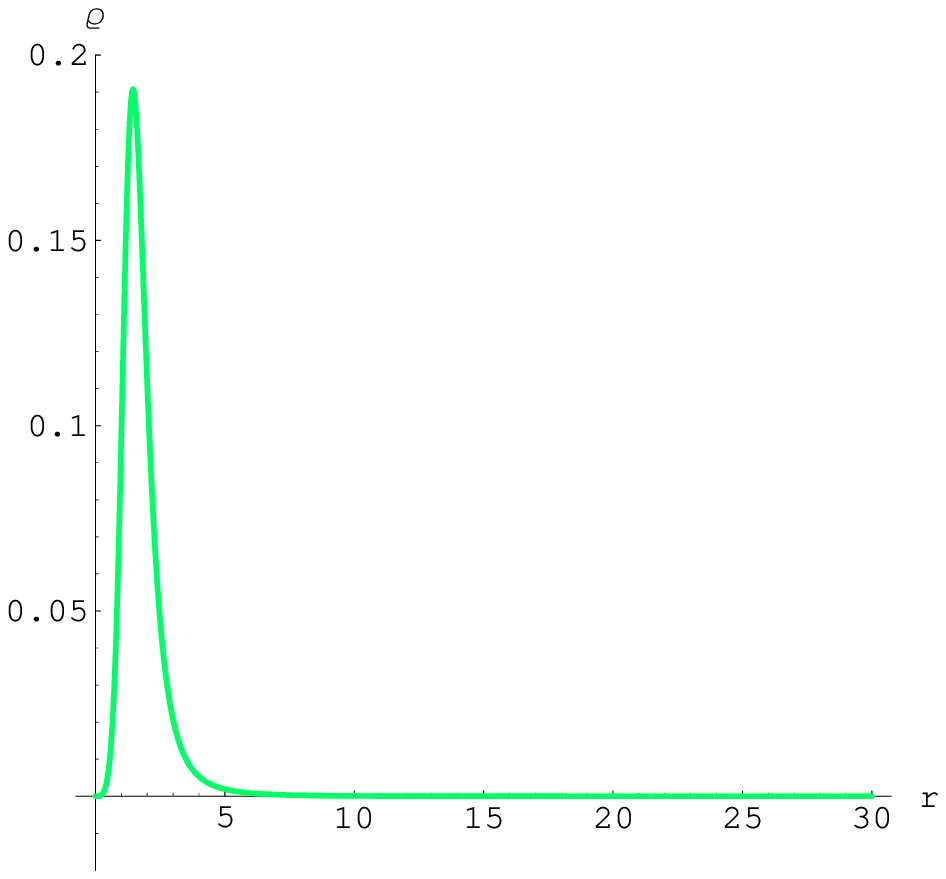} \hspace{1cm}
\includegraphics[height=3cm]{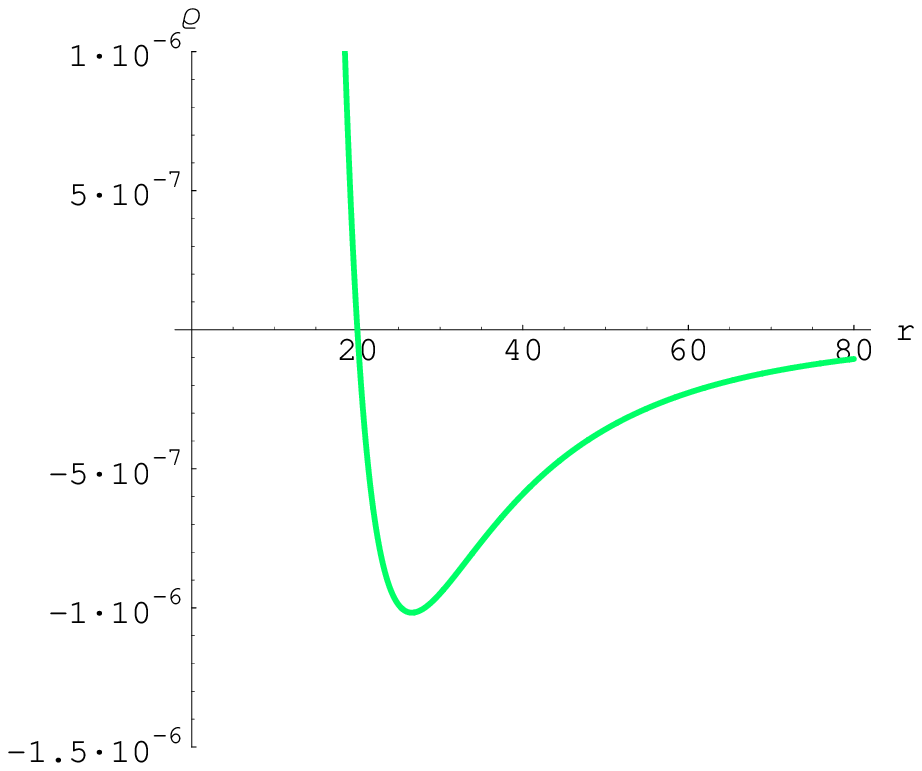} 
\caption{\label{rho} 
Plot of the effective energy density $\rho$ 
 for $m = 10$ and $\gamma \delta \sim 1$.
 The plots represent the energy density (in the center plot) and two zooms
 for $r \sim 0$ and $r \gtrsim 2m$.
}
  \includegraphics[height=3cm]{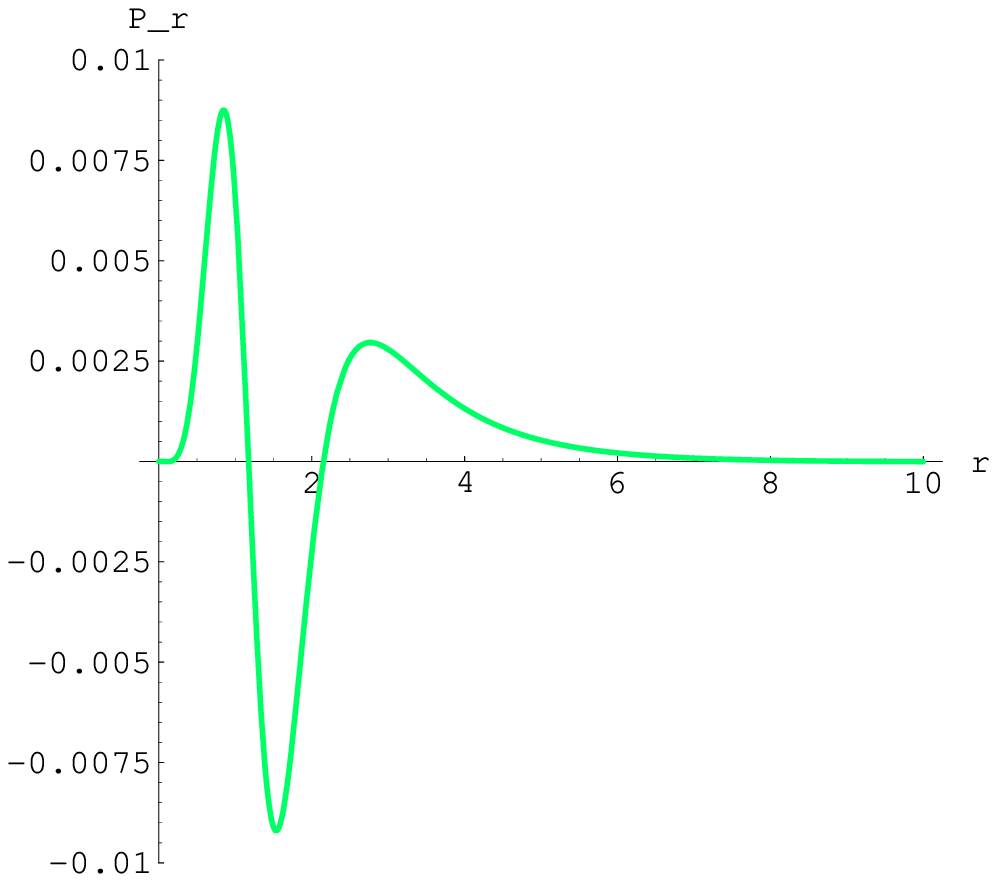} \hspace{2cm}
  \includegraphics[height=3cm]{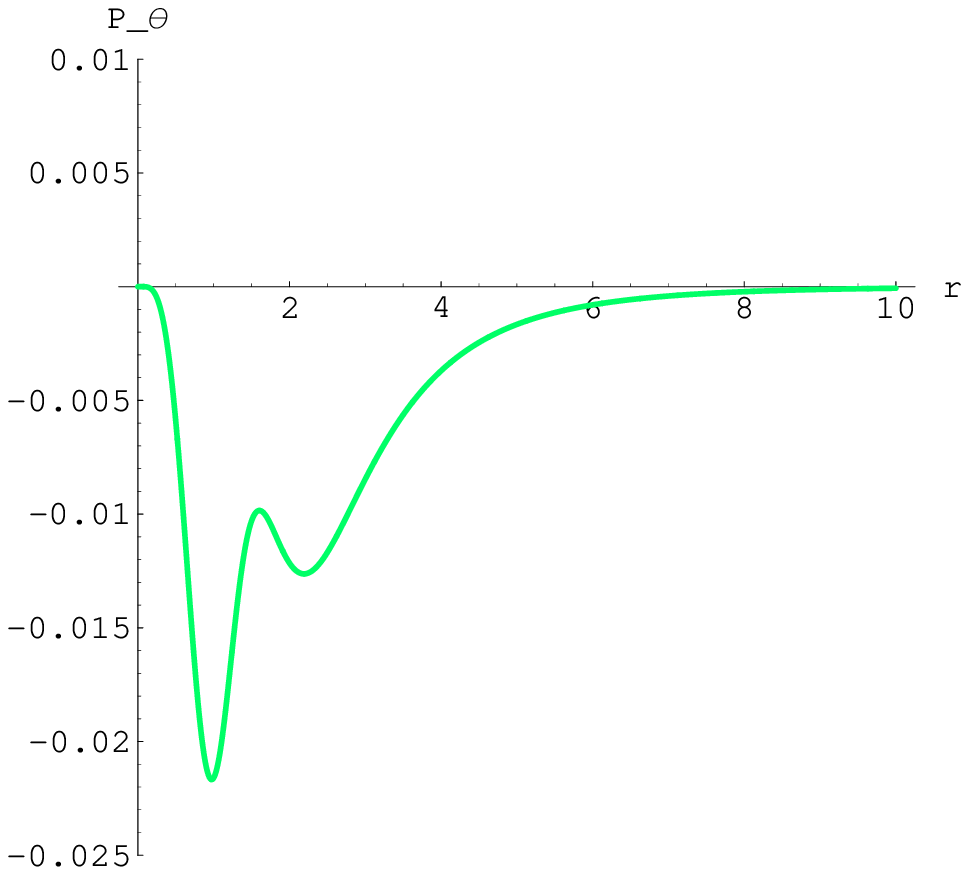} \hspace{1cm}
\includegraphics[height=3cm]{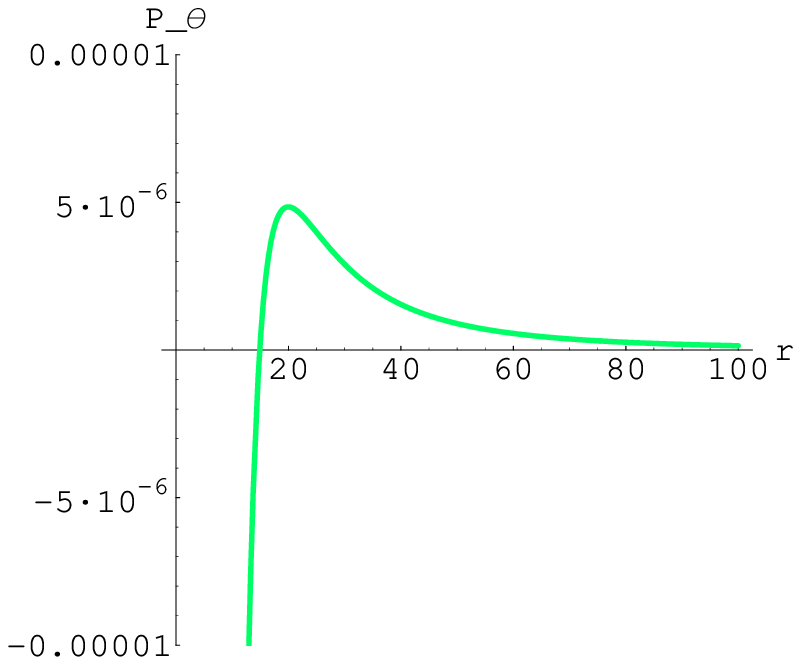}
  \end{center}
   \caption{\label{prpt} 
 Plot of the pressures $P_r$, $P_{\theta}$
 for $m = 10$ and $\gamma \delta \sim 1$.
 The first plot represents the pressure $P_r$,
 the second and the third plot represent the pressure $P_{\theta}$
 for $r \leqslant 2m$ and $r \gtrsim 2m$
  respectively.
 }
  \end{figure}
\noindent
We can introduce the null coordinate $v$ to express the metric 
(\ref{generalmet}) in the Bardeen form.
The null coordinate $v$ is defined by the relation $v = t + r^{\ast}$, where 
$r^{\ast} = \int^r d r/\sqrt{f (r) g(r)}$ and the 
differential 
is $d v = d t + d r/\sqrt{f(r) g(r)}$. In the new coordinate the metric is
\begin{eqnarray}
 d s^2 = - g(r) d v^2 + 2 \sqrt{\frac{g(r)}{f(r)}} \, d r  dv + h^2(r) (d \theta^2 + \sin^2 \theta d \phi^2).
 \label{Bardeen}
\end{eqnarray}
\noindent
We can interpret our black hole solution has been generated by an effective 
matter fluid that simulates the loop quantum gravity corrections (in analogy with
the paper \cite{BR}).
The effective gravity-matter system satisfies by definition of the Einstein 
equation ${\mathbf G} =8 \pi {\mathbf T}$, where $\mathbf T$ is the 
effective energy tensor.
In this paper we are not interested to the explicit form of the stress energy
tensor however we give a plot of the tensor components for completeness.
It is possible to calculate the stress energy tensor by using the Einstein 
equations. The stress energy tensor for a perfect fluid compatible with the space-time
symmetries is $T^{\mu}_{\nu} = (- \rho, P_r, P_{\theta}, P_{\theta})$
and in terms of the Einstein tensor the components are 
$\rho= - G^t_t/8 \pi G_N$, $P_r = G^r_r/8 \pi G_N$
and $P_{\theta}= G^{\theta}_{\theta}/8 \pi G_N$.
We can calculate the Einstein tensor components using the metric in the 
table of section one. The components of the effective energy 
tensor that simulates quantum gravity effects are plotted in Fig.\ref{rho} and Fig.\ref{prpt}. 
In the plots we have amplified quantum gravity effects taking $\gamma \delta \sim 1$, 
however it is evident that the energy density and pressure are meaningful
only in the Planck region. In the contrary for $r \gg l_P$, energy density and pressure 
tend to zero. To the second order in $\delta^2$ the energy density is 
\begin{eqnarray}
\rho = \frac{m(7 m^2 - r^4 + m r (2 r^2 -3)) \gamma^2 \delta^2}{ 8 \pi G_N \, r^7}.
\label{densityserie}
\end{eqnarray}
If we develop the semiclassical metric solution of section one
to order $\delta^2$ and we introduce the result 
in the Einstein tensor $G^{\mu}_ {\nu}$,
we obtain (to order $\delta^2$) the energy density (\ref{densityserie}).
The semiclassical metric to zero order in $\delta$ is the classical Schwarzschild solution ($g_{\mu \nu}^{C}$) 
and $G^{\mu}_{\nu}(g^C) \equiv 0$.

\begin{figure}
 \begin{center}
  \includegraphics[height=6cm]{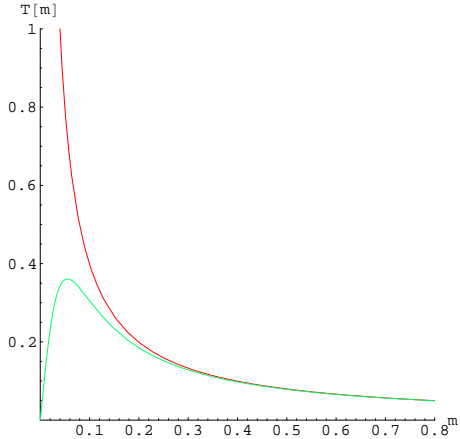}
  \end{center}
  \caption{\label{TBH} 
 Plot of the temperature as function of the mass 
 $m$ and $\gamma \delta \sim 1$ to amplify the quantum gravity effects; 
 the red trajectory corresponds to the Hawking temperature 
 $T_{H} = 1/8 \pi m$ and the green trajectory corresponds to the quantum geometry 
 temperature.}
  \end{figure}

\paragraph{Temperature.}
In this paragraph we are interested in calculate the temperature and 
entropy for our modified black hole solution and analyze the evaporation process.
The Bekenstein-Hawking temperature is given in terms of the surface gravity
$\kappa$ by $T_{BH}= \kappa/2 \pi$. The surface gravity is defined by
\begin{eqnarray}
\kappa^2 = - \frac{1}{2} g^{\mu \nu} g_{\rho \sigma} \nabla_{\mu} \chi^{\rho} \nabla_{\nu}
\chi^{\sigma} = - \frac{1}{2} g^{\mu \nu} g_{\rho \sigma}  \Gamma^{\rho}_{\mu 0} \Gamma^{\sigma}_{\nu 0},
\end{eqnarray}
where $\chi^{\mu}=(1,0,0,0)$ is a timelike Killing vector and $\Gamma^{\mu}_{\nu \rho}$
is the connection compatibles with the metric $g_{\mu \nu}$ of (\ref{generalmet}).
Using the semiclassical metric in the table of section one we can calculate the surface gravity
in $r = 2m$ obtaining 
$\kappa^2 = - \frac{1}{4} g^{00} g^{11} \left(\frac{\partial g_{00}}{\partial r} \right)^2 =
\left(\frac{16 m}{64 m^2 + \gamma^2 \delta^2} \right)^2,
\label{surface}$
therefore the temperature is 
\begin{eqnarray}
T_{BH} = \frac{8 m}{\pi (64 m^2 + \gamma^2 \delta^2)}.
\label{Temperatura}
\end{eqnarray}
The temperature in (\ref{Temperatura}) coincides with the Hawking temperature 
in the limit $\delta \rightarrow 0$. 
In Fig.\ref{TBH} we have a plot of the temperature as a function of
the black hole mass $m$. The red trajectory corresponds to the Hawking 
temperature and the green trajectory corresponds to the semiclassical one.
There is a substantial difference for small values of the mass, in fact 
the semiclassical temperature tends to zero and does not diverge for $m\rightarrow 0$.
The temperature is maximum for $m = \gamma \delta/8$ and $T_{\rm max}=1/2 \pi \gamma \delta$.
\paragraph{Entropy.} Another interesting quantity to calculate is the entropy and its 
quantum corrections. By definition $S_{BH}=\int dm/T_{BH}(m)$ and we 
obtain 
\begin{eqnarray}
S_{BH} = 4 \pi m^2 + \frac{\gamma^2 \delta^2}{16} \ln m^2.
\label{entropym}
\end{eqnarray}
On the other hand the event horizon area (in $r=2m$) is 
\begin{eqnarray}
A = \int d \phi d \theta \sin \theta \, Y^2 (r) = 16 \pi m^2 +  \frac{ \pi \gamma^2 \delta^2}{4}.
\label{area}
\end{eqnarray}
Using (\ref{area}) we can express the entropy in terms of the event horizon
area 
\begin{eqnarray}
&& S_{BH} = \frac{A}{4}  + \frac{\gamma^2 \delta^2}{16} \ln\left[\frac{A}{4} \left(1 - \frac{\pi \gamma^2 \delta^2}{4 A} \right) \right] - \frac{\pi \gamma^2 \delta^2}{16} - \frac{\gamma^2 \delta^2}{16} \ln 4 \pi =
\nonumber \\
&&  \hspace{0.8cm} = \frac{A}{4}  + \frac{\gamma^2 \delta^2}{16} \ln\left(\frac{A}{4}\right) 
+  \frac{\gamma^2 \delta^2}{16}  \ln \left(1 - \frac{\pi \gamma^2 \delta^2}{4 A} \right)
 - \frac{\pi \gamma^2 \delta^2}{16} - \frac{\gamma^2 \delta^2}{16} \ln 4 \pi =
 \nonumber \\
&&  \hspace{0.8cm} =\frac{A}{4}  + \frac{\gamma^2 \delta^2}{16} \ln\left(\frac{A}{4}\right) 
-  \frac{\gamma^2 \delta^2}{16}  \sum_{n=1}^{\infty} \frac{1}{n} \left(\frac{\pi \gamma^2 \delta^2}{4A} \right)^n
 - \frac{\pi \gamma^2 \delta^2}{16} - \frac{\gamma^2 \delta^2}{16} \ln 4 \pi.
\label{entropyarea}
\end{eqnarray}
In the last step we have developed the second $\ln$-function for $\delta^2 \ll A$ 
to compare our result with the general formula in letterature $S = \frac{A}{4}  + \rho 
\ln\left(\frac{A}{4}\right) + \sum_{n=1} c_n\left(\frac{4}{A}\right)^n + const.$. 
Our calculation reproduces the standard area term and the logarithmic correction. 
In (\ref{entropyarea}) the $(\delta^2/A)^n$ corrections appear in the form of 
another logarithmic function.
  
\begin{figure}
 \begin{center}
  \includegraphics[height=6cm]{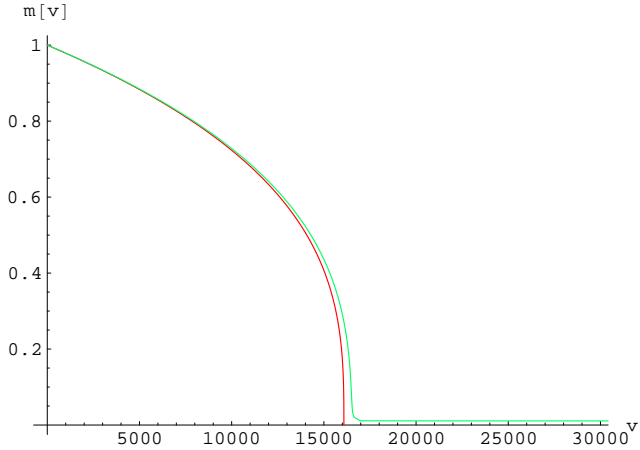}
  \end{center}
  \caption{\label{Plotmv} 
   Plot of $m(v)$ 
 for $m_0 = 1$, $\gamma \delta \sim 1$ and $\forall v \geqslant 0$.
 In the plot it is evident that classically (red trajectory) the mass evaporate in a finite time,
 on the contrary using the solution suggested by loop quantum black hole the mass
 evaporate in an infinite time. }
  \end{figure}

\paragraph{Evaporation process.}In the previews paragraph the metric (\ref{Bardeen}) is
a static solution of the effective Einstein equation of motion 
outside the event horizon. That solution is a right even 
if the mass is a function of the null coordinate $v$ but with 
a non static effective stress energy tensor. In this paper 
we are not interested in the explicit form of the energy tensor.
Instead we are interested in the evaporation process 
of the mass and in particular in the energy flux from the 
black hole. The luminosity can be estimated using the 
Stefan law and it is given by ${\mathcal L}(m)= \sigma A(m) T_{BH}^2(m)$,
where (for a single massless field with two degree of freedom)  
$\sigma = \pi^2/60$, $A(m)$ is the event horizon area and $T_{BH}(m)$
is the Bekenstein-Hawking temperature calculated in the previous section. 
At the first order in the luminosity the metric (\ref{Bardeen}) witch incorporates 
the decreasing mass as function of the null coordinate $v$ is also a solution
but with a new effective stress energy tensor as underlined previously. 
Introducing the results (\ref{Temperatura}) and (\ref{area}) of the previous paragraph  
in the luminosity $\mathcal{L}$ 
we obtain 
\begin{eqnarray}
\mathcal{L}(m) = \frac{2^{16} m^6 + 2^{10} \gamma^2 \delta^2 m^4}{60 \pi (64 m^2 + \gamma^2 \delta^2)^4}.
\label{lumini}
\end{eqnarray}
Using (\ref{lumini}) we can solve the fist order differential equation
\begin{eqnarray}
- \frac{d m(v)}{d v} = \mathcal{L}[m(v)]
\label{flux}
\end{eqnarray}
to obtain the mass function $m(v)$. The result of integration with initial 
condition $m(v = 0) = m_0$ is 
\begin{eqnarray}
5120 \pi (m- m_0)^3 + 720 \pi \gamma^2 \delta^2 (m- m_0) - \frac{45 \pi \gamma^4 \delta^4}{4} 
\left(\frac{1}{m} - \frac{1}{m_0}\right)
- \frac{5 \pi \gamma^6 \delta^6}{256}\left(\frac{1}{m^3} - \frac{1}{m_0^3}\right) = - v.
\label{v(m)}
\end{eqnarray} 
In Fig.\ref{Plotmv} there is an implicit plot of $m(v)$ and it is 
evident the difference with the classical result. Classically (red trajectory)
the mass evaporates in a finite time but at the semiclassical level (green trajectory)
the mass  evaporates in an infinite time.
We can calculate the value of $m$ where the concavity of $m(v)$ changes. From 
the second derivative of the function $m(v)$ we obtain 
\begin{eqnarray}
\frac{d^2 m(v)}{d v^2} \sim - \mathcal{L}[m(v)] \frac{1024 m^3 (32 m^2 - \gamma^2 \delta^2)}{15 \pi(64 m^2 + \gamma^2 \delta^2)^4}.
\label{secderiv}
\end{eqnarray}
and equalling (\ref{secderiv}) to zero we obtain the mass $m_c = \frac{\gamma \delta}{4\sqrt{2}}$.
The value $m_c$ is in the order Planck mass and at this scale it is inevitable a complete quantum analysis of the problem. However in this semiclassical study the evaporation process 
needs of infinite time.

\section*{Conclusions}
In this paper we have 
extended to all space time the regular solution calculated
and studied inside the event horizon in the preview paper \cite{SS}.
The solution has been obtained 
solving the Hamilton equation of motion 
for the Kantowski-Sachs space-time \cite{KS} using the regularized Hamiltonian 
constraint suggested by loop quantum gravity.
The semiclassical solution reproduces the Schwarzschild 
solution at large distance from the event horizon 
but it is substantially different 
in the Planck region near the point $r=0$, where the singularity is 
(classically) localized. 
The solution has two event horizons :
the first in $r=2m$ and the other near the point $r=0$
(this suggests a similarity  with the result in ``asymptotic safety quantum gravity" \cite{BR},
but the radius of such horizon is smaller than the Planck length  
and in this region it is inevitable a complete quantum analysis 
of the problem \cite{ABM}).

In this paper we have concentrated our attention on the evaporation process
and we have calculated the temperature, entropy ( with all the correction suggested by 
the particular model) and the mass variation formula as seen by a distant observer at 
the time $v$. The main results are:
\begin{enumerate}
\item 
The classical black hole singularity near $r \sim 0$
disappears from the semiclassical solution.  
The classical divergent curvature invariant is bounded in the
the semiclassical theory and in particular 
$\rm R_{\mu \nu \rho \sigma} \, \rm R^{\mu \nu \rho \sigma} \rightarrow 0$
for $r \rightarrow 0$ and for $r \rightarrow \infty$.
\item The Bekenstein-Hawking temperature $T_{BH}(m)$ is regular
for $m \sim 0$ and tends to zero
\begin{eqnarray}
T_{BH} = \frac{8 m}{\pi (64 m^2 + \gamma^2 \delta^2)}.
\end{eqnarray}
\item  The black hole entropy in terms of the horizon area 
reproduces the $A/4l_P^2$ term but contain also the $\ln$-correction and
all the other correction in $(l_P^2/A)^n$
\begin{eqnarray}
 S=\frac{A}{4 l_P^2}  + \frac{\gamma^2}{16} \ln\left(\frac{A}{4 l_P^2}\right) 
+  \frac{\gamma^2}{16}  \ln \left(1 - \frac{4 \pi \gamma^2 l_P^2}{16 A} \right)
 + {\rm const}, 
 \end{eqnarray}
(where we have repristed the length units). 
\item The evaporation process needs an infinite time in our semiclassical analysis  
but the difference with the classical result is evident only at the Planck scale 
when the black hole mass is the order $m \sim m_c = \gamma \delta/4 \sqrt{2}$.
In this extreme energy conditions it is inevitable a complete quantum gravity 
analysis that can implies a complete evaporation \cite{AB}.
\end{enumerate}

We think that the semiclassical analysis performed here will sheds light on 
the problem of the ``information loss" in the process of black 
hole formation and evaporation
but a complete quantum analysis is necessary to understand 
what happen in the Planck region. See in particular \cite{AB} for a possible
physical interpretation of the black hole information loss problem.



\section*{Acknowledgements}

We are strongly indebted to Roberto Balbinot 
for crucial criticisms, inputs and suggestions.
We are grateful also to Alfio Boananno, Eugenio Bianchi and Guido Cossu 
for many important and clarifying discussion.



\end{document}